\documentclass[aps,pra, twocolumn,groupaddress,twocolumn,10pt]{revtex4-1} 
\usepackage[utf8]{inputenc}

\usepackage{amsmath}
\usepackage{amssymb}
\usepackage{amsfonts}
\usepackage{braket}
\usepackage{latexsym}
\usepackage{paralist}
\usepackage{nicefrac}
\usepackage{dsfont}
\usepackage{colortbl}

\usepackage{graphicx}

\usepackage{nicefrac}
\usepackage{xcolor}
\usepackage{subfigure}
\usepackage{psfrag}
\usepackage{tikz}

\usepackage{natbib}
\usepackage{fancyhdr}
\usepackage[colorlinks=true,citecolor=blue,linkcolor=blue]{hyperref}

\usepackage{ulem} 
\usepackage{hhline}
\usepackage{multirow}
\hypersetup{
    bookmarks=true,         
    bookmarksopen=true,			
    bookmarksopenlevel=1,		
    unicode=false,          
    pdftoolbar=true,        
    pdfmenubar=true,        
    pdffitwindow=false,     
    pdfstartview={FitH},    
    pdftitle={High-fidelity Rydberg-blockade entangling gate using shaped, analytic pulses},    
    pdfauthor={Lukas Theis},     
    urlcolor=blue           
}

	\newcommand{\comut}[2]{\left[ #1 , #2 \right]} 
	\newcommand{\op}[1]{\hat{\mathrm{#1}}}
	\newcommand{\id}{\mathds{\hat{1}}}
	
	\newcommand{\dt}{\mathrm{d}}

	\renewcommand{\cos}[1]{\mathrm{cos}\LR{#1}}
	
	\renewcommand{\exp}[1]{\mathrm{exp}\LR{#1}}

	\newcommand{\del}{\mathrm{\Delta}}
	\newcommand{\ketbra}[2]{\ket{#1}\!\bra{#2}}
	
	\newcommand{\e}{\varepsilon}
	\newcommand{\LR}[1]{\left(#1\right)}

	\newcommand{\PTrace}[2]{\text{Tr}_{#1}\left\{#2\right\}}	

	\newcommand{\reffig}[1]{Fig.\,\ref{fig:#1}}
	\newcommand{\refeq}[1]{(\ref{eq:#1})}
	\newcommand{\reftab}[1]{Tab.\,\ref{tab:#1}}

\makeindex

\begin{document}

\title{High-fidelity Rydberg-blockade entangling gate using shaped, analytic pulses}

\author{L. S. Theis}\email{luk@lusi.uni-sb.de}
\author{F. Motzoi}
\author{F. K. Wilhelm}
\affiliation{Theoretical Physics, Saarland University, 66123 Saarbr{\"u}cken, Germany}
\author{M. Saffman}
\affiliation{Department of Physics, 1150 University Avenue, University of Wisconsin-Madison, Madison, Wisconsin 53706, USA}

\makeatletter
\def\Dated@name{Date: }
\makeatother
\date{\today}

\begin{abstract}
We show that the use of shaped pulses improves the fidelity of a Rydberg blockade two-qubit entangling  gate by several orders of magnitude compared to previous protocols based on square pulses or optimal control pulses.  Using analytical Derivative Removal by Adiabatic Gate (DRAG) pulses that reduce excitation of primary leakage states  and an analytical method of finding the optimal Rydberg blockade we generate  Bell states with a fidelity of $F>0.9999$ in a 300 K environment  for a gate time of only $50\;{\rm ns}$, which is an order of magnitude faster than previous protocols. These results establish the potential of neutral atom qubits with Rydberg blockade gates for scalable quantum computation.  
\end{abstract}

\maketitle

\section{Introduction}
The Rydberg blockade mechanism  introduced in \cite{Jaksch2000} has been demonstrated to be capable of creating bipartite entanglement with fidelity of $\sim 0.7 - 0.8$\cite{Maller_PRA_92_022336,Jau2016}. There is good reason to believe that the fidelity achieved to date is not a fundamental limit, but is due to experimental perturbations and the high sensitivity of Rydberg states to external fields \cite{Saffman2016}. With the expectation that experimental techniques will continue to improve it is important to address the question of the intrinsic fidelity limit of the Rydberg blockade gate. Detailed analysis with constant amplitude Rydberg excitation pulses revealed a Bell state fidelity limit of $F_B\sim 0.999$  in Rb or Cs atoms in a 300 K environment\cite{Zhang_PRA_85_042310}. Other work has sought to improve on this with optimal control pulse shapes\cite{Muller2011,Goerz_PRA_90_032329}, adiabatic excitation\cite{Muller_PRA_89_032334,Rao2014}, or simplified protocols that use a single Rydberg pulse\cite{Han2016,Su2016}. However, none of the analyses to date that consistently account for Rydberg decay and excitation leakage to neighboring Rydberg states have provided a fidelity better than 0.999. This leaves open the question of whether or not the Rydberg gate will be capable of reaching the $0.9999$ level or better that appears necessary for scalable quantum computation with a realisitc overhead in terms of qubit numbers for logical encoding\cite{Devitt2013}.

In this work we show that Rydberg gates with $F_B>0.9999$ are possible with Cs atoms in a  300 K environment and $F_B>0.99999$ in a 4K environment. This advance is made possible using simple and smooth analytic shaped pulses that are designed to suppress leakage at a discrete set of frequencies\cite{Motzoi_PRL_103_110501} corresponding to neighboring Rydberg states. By suppressing the leakage orders of magnitude more effectively than is possible with square, or simple Gaussian pulses, we are able to run the gate at least an order of magnitude faster than previous protocols, which is fast enough to keep the spontaneous emission error low and achieve high fidelity. Drastically reducing the gate time also has advantages in the short term, by avoiding the onset of other experimental errors that increase with time, such as technical noise. We find a gate time close to 50 ns, which is fast enough to be competitive with superconducting qubits while retaining much longer coherence times \cite{Egger_SUST_2013,Ghosh_PRA_87_022309}.
\begin{figure}
 \includegraphics[width=.98\linewidth]{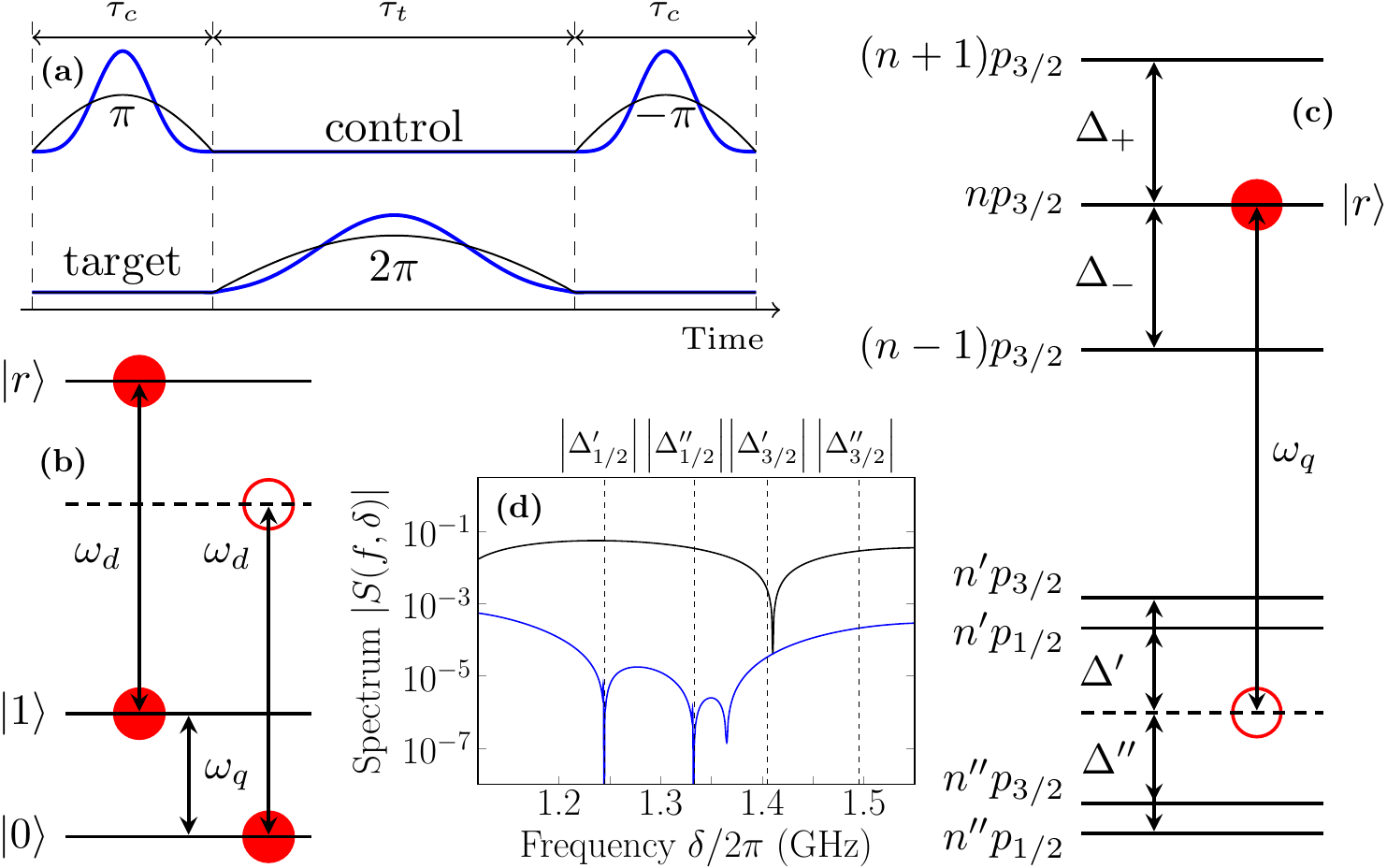}
 \caption{\label{fig:sequence}(color online) a) DRAG pulse sequence (blue) and initial Gaussian waveform, (thin black)  to implement a two-qubit entangling gate. Pulse durations are $\tau_c, \tau_t$ for the control and target atoms. The control amplitudes are shown on the same scale. b) Level diagram for one-photon Rydberg excitation with laser frequency $\omega_d$. c) Detail of the Rydberg level structure and detunings. d) Spectrum of pulse on control atom for Gaussian (black) and DRAG (blue) waveforms.}
\end{figure}

\section{Rydberg excitation} The free evolution and gate Hamiltonians $\op{H}_d$ and $\op{H}_g$, respectively, of a single Rydberg atom in its lab frame, are given by ($\hbar=1$ everywhere)
\begin{subequations}\label{eq:ham_lab}\begin{align}
	\op{H}_{d} & = \omega_g\ketbra{g}{g}+\omega_q\ketbra{1}{1} + \sum\limits_{r'}\omega_{r'}\ketbra{r'}{r'}\\
	\op{H}_{g} & = \Omega(t)\sum\limits_{r'}\left(\frac{n}{n'}\right)^{3/2}\left(\ketbra{r'}{0}+\ketbra{r'}{1}\right) + \text{h.c.}
\end{align}\end{subequations}
whereby $\ket{g}$ denotes some auxiliary level we will use to model decay. Here $r'$ is shorthand for the set of quantum numbers specifying the Rydberg states and $n,n'$ are the principal quantum numbers. The matrix elements and the Rabi coupling for single photon excitation to high lying Rydberg states scale as $1/n^{3/2}$. For Cs, the ground hyperfine splitting is $\omega_q/2\pi=9.1926\;\rm{GHz}$. The set of states $\{\ket{r'}\}$ describes all relevant Rydberg states. We assume there is negligible coupling of any of the states to $\ket{g}$ due to the control $\Omega(t)$, hence without loss of generality we set its energy to zero, i.e. $\omega_g=0$. The control field has in-phase control only,
\begin{align}\label{eq:control_field}
	\Omega(t) & = \e_x(t)\cos{\omega_d t}.
\end{align}
Usually, atoms are driven on resonance with the $\ket{1}\leftrightarrow\ket{r}$ transition, so that $\omega_d=\omega_r-\omega_1$ with $\omega_r$ being the frequency of the target Rydberg state $\ket{r}$. In order to remove any oscillation on the order of $\omega_d$ from the dynamics, we choose to work in a frame rotating with $\omega_d$ in the remainder of this work. The pulse sequence which we will use to implement a two-qubit entangling gate is illustrated in \reffig{sequence}a). When control and target atoms are initially prepared in their $\ket{1}$ state, the desired Rydberg state $\ket{r}$ of the target atom will be Rydberg-blockaded by ${\sf B}_0$ during the $2\pi$-pulse due to the control atom's $\ket{r}$ state being populated. Hence, the $2\pi$-pulse will ideally produce a phase shift of $\pi$ on the state $\ket{1}$ of the target atom. This scheme implements an entangling C$_{\rm Z}$ gate\cite{Jaksch2000},  $\op{U}_{\rm C_Z} = \mathrm{diag}(1,-1,1,1)$ in the computational basis $\{\ket{00},\ket{01},\ket{10},\ket{11}\}$. This differs from the phase gate matrix of \cite{Jaksch2000} due to our use of $-\pi$ instead of $\pi$ for the last pulse which results in slightly better gate fidelity.

The Hamiltonian of the compound system, control and target atom, can be written as 
\begin{align}\label{eq:ham_tot}
  \op{H} & = \op{H}_{\rm control} \otimes \id + \id \otimes \op{H}_{\rm target} + \sum\limits_{i,j}{\sf B}_{r_i,r_j}\ketbra{r_i,r_j}{r_i,r_j}.
\end{align}
Here, the ${\sf B}_{r_i,r_j}$ quantify the Rydberg interaction strength between all relevant Rydberg states $\ket{r_i}$ of the control atom and $\ket{r_j}$ of the target atom, including all possible leakage levels  depicted in \reffig{sequence}c). The desired excitation is resonant between $\ket{1}$ and $\ket{r}$, with leakage channels to both $(n\pm 1)p_{3/2}$ states (detunings $\del_{\pm}$).

To remove  leakage to $np_{1/2}$ states we assume a specific implementation in Cs atoms where qubit state $\ket{1}$ is mapped to $\ket{1'}=\ket{f=4,m=4}$ before and after the Rydberg gate. With $\sigma_+$ polarized excitation light $\ket{1'}$ only couples to states $\ket{np_{3/2},f=5,m_f=5}$ so there is no leakage to $np_{1/2}$ states, and errors due to coupling to multiple hyperfine states within the $np_{3/2}$ levels are also suppressed. For compactness of notation we refer to $\ket{1'}$ as $\ket{1}$ in the following.

In addition to leakage to the blockaded target Rydberg state during the $2\pi$ pulse, significant leakage channels exist for the $\pi$ pulse when the control qubit is initially in the $\ket{0}$ state (see \reffig{sequence}c).  The $\ket{0}$ state is coupled to Rydberg states $\ket{n'p_{1/2,3/2}}$ and $\ket{n''p_{1/2,3/2}}$. The Cs $6s_{1/2}-np_{1/2}$ oscillator strength is anomalously small, as was first explained by Fermi\cite{Fermi1930}, and for the states S1, S2 of primary interest in \reftab{params} we estimate the ratio of Rabi coupling strengths to $np_{1/2}$ states as compared to $np_{3/2}$ states as $<1/300$\cite{Lorenzen1978}. The leakage to $np_{1/2}$ states in Cs with Gaussian pulses is therefore negligible. Nevertheless we have still included possible leakage to $np_{1/2}$ states in order to substantiate the generality of our approach. The detunings for these transitions to $np_{1/2,3/2}$ states are $\del',\del''$.  In what follows we will refer to the interaction between two target Rydberg states $\ket{np_{3/2}}$ as ${\sf B}_0$.

\section{Design of DRAG pulses} An analytic tool to minimize leakage errors is the \emph{Derivative Removal by Adiabatic Gate} (DRAG) method \cite{Motzoi_PRL_103_110501} which is based on shaping both in- and out-of-phase control of the system. The method has been further developed \cite{Motzoi_PRA_88_062318} and provides a general toolbox to design frequency-selective pulses, a form of counter-diabatic driving \cite{Demirplak_Chem_107_9937,Torrontegui_shortcuts}. Under the assumption -- which can be derived from a Magnus expansion \cite{Warren_JChemPhys_81_5437} in the interaction picture -- that the finite Fourier transform
\begin{align}\label{eq:fourier}
      S(f,\delta)=\int\limits_0^{T}\dt t\;f(t)e^{i\delta t}
\end{align}
gives a good first-order estimate of the evolution, DRAG pulses can be alternately and more simply be derived so that they have no spectral power at certain frequencies $\{\delta_j\}$ with $j=1,\ldots,m$ . In contrast to previous work on DRAG controls, we will utilize only a shaped in-phase control $\e_x(t)$ and no additional out-of-phase quadrature, which has been suggested in earlier work \cite{Gambetta_PRA_83_012308}. This simplifies the experimental implementation. A key requirement for the method to work is that the first $N=2m$ derivatives of a pulse $f(t)$ vanish at its beginning (0) and end (T), so that we can use integration by parts to show that
\begin{align}\label{eq:partint}
	S(f,\delta) = \left(\frac{i}{\delta}\right)^N\int\limits_0^{T}\dt t\;\frac{\dt^N f(t)}{\dt t^N}e^{i\delta t}.
\end{align}
To obtain a control shape $\e_x(t)$ which satisfies $S(\e_x,\delta_j)=0$ for all $j=1,\ldots,m$ we make the expansion
\begin{align}\label{eq:ansatzDrag}
	\e_x(t) = \e_x^{(0)}(t) + \sum\limits_{k=1}^{N/2}\alpha_{2k}\frac{\dt^{2k} \e_x^{(0)}(t)}{\dt t^{2k}},
\end{align}
whereby $\e_x^{(0)}(t)$ is some smooth initial shape, e.g. a Gaussian pulse, which satisfies  Eq.\refeq{partint}. This particular ansatz as an expansion in terms of derivatives is motivated by a sequence of adiabatic transformations that yield instantaneous-time control and aid analytical solutions to the dynamics. Demanding $S(\e_x,\delta_j)=0$ for all $j=1,\ldots m$ and utilizing Eq.\refeq{partint} as well as Eq.\refeq{ansatzDrag} leads to a system of $m$ equations for the coefficients $\alpha_k$,
\begin{align}
	1+\sum\limits_{k=1}^{N/2}\alpha_{2k}\left(-i\delta_j\right)^{2k} = 0,\quad j=1,\ldots,m.
\end{align}
For instance, if two leakage transitions at $\delta_1$ and $\delta_2$ need to be suppressed, the corresponding real-valued solutions for the coefficients $\alpha_k$ in Eq.\refeq{ansatzDrag} read
$\alpha_2 = -\left(\frac{1}{(\delta_1)^2}+\frac{1}{(\delta_2)^2}\right)$, $\alpha_4 = \frac{1}{(\delta_1)^2(\delta_2)^2}$. It is important to note that the solutions presented here minimize the error at every instant of time as can be seen from a rigorous iterative application of adiabatic transformations which essentially arrives at the same result \cite{Motzoi_PRA_88_062318}. This substantiates that the ansatz in Eq.\refeq{ansatzDrag} is preferable over other possible waveforms $f(t)$ that solely satisfy $S(f,\delta_j)=0$.

\section{Gate analysis}
\subsection{Population error}
We proceed to demonstrate how Gaussian pulses with DRAG components help to improve over previous methods by several orders of magnitude. Since the main advantage of Gaussian and DRAG shapes is an exponential suppression of leakage, we first focus on population error arising from leakage channels to other Rydberg states as shown in \reffig{sequence}c). For our simulations we use the system parameters that are listed in \reftab{params}. The two different settings, S1 and S2, respectively, belong to two possible one-photon-excitation schemes starting from the Cs $6s_{1/2}$ state. Leakage errors are expected to be worse in S2 due to smaller energy splittings at higher Rydberg states, whilst lifetimes in S2 are better by roughly a factor of two. As initial pulses for DRAG and Gaussian control, we utilize generalized Gaussians of duration $T$
\begin{align}
 \e_G(t) & = A_{\theta}\left[\exp{-\frac{(t-T/2)^2}{2\sigma^2}}-\exp{-\frac{(T/2)^2}{2\sigma^2}}\right]^{N+1}
\end{align}
with a standard deviation $\sigma=2T/3$ and a pulse  area  $\theta$ determined by the value of $A_{\theta}$\cite{alphanote}. The exponent $N+1$ ensures that the first $N=2m$ derivatives of the Gaussian vanish at times $t=0$ and $t=T$. Note that $N$ here is the same as e.g. in \refeq{ansatzDrag}, so that we meet the conditions for Eq.\refeq{partint} to hold. Unless stated otherwise, we fix the pulse length $\tau_c$ for the $\pm\pi$-pulses on the control atom to $\tau_c=\tau_t/2$.   

\begin{table}[]
	\centering
	\label{my-label}
	\begin{tabular}{|c|c|c||c|cc|}
	\hline
	\multicolumn{1}{|c|}{\multirow{2}{*}{Parameter}} & \multicolumn{2}{c||}{Value} & \multirow{2}{*}{Parameter} & \multicolumn{2}{c|}{Value}       \\
	\multicolumn{1}{|c|}{}                           & S1            & S2         &                            & \multicolumn{1}{c|}{S1}    & S2 \\ \hline\hline
	$n$                                             & 107           & 141        &  $\tau_n\;(\mu s)$         & \multicolumn{1}{c|}{538}   & 969 \\
	$n'$                                            & 106           & 138        &  $\del_+/2\pi\;({\rm GHz})$     & \multicolumn{1}{c|}{-5.534} & -2.507   \\
	$n''$                                           & 105           & 137        &  $\del_-/2\pi\;({\rm GHz})$     & \multicolumn{1}{c|}{5.694} & 2.562   \\
	$\del'_{1/2}/2\pi\;({\rm GHz})$			& -2.961         & -1.245      &  $\del'_{3/2}/2\pi\;({\rm GHz})$& \multicolumn{1}{c|}{-3.161} & -1.333   \\
	$\del''_{1/2}/2\pi\;({\rm GHz})$			& 3.256         & 1.495      &  $\del''_{3/2}/2\pi\;({\rm GHz})$& \multicolumn{1}{c|}{3.051} & 1.405 \\
	${\sf B}_0/2\pi\;({\rm GHz})$ 				& 1.54		& 0.68	     & 	$b_{n,n}$ & \multicolumn{2}{c|}{1}\\
	$b_{n,n'}$ & \multicolumn{2}{c||}{0.85} & $b_{n,n''}$ & \multicolumn{2}{c|}{0.80}\\
	$b_{n,n+1}$ & \multicolumn{2}{c||}{1.02} & $b_{n,n-1}$ & \multicolumn{2}{c|}{0.97}
	\\\hline
	\end{tabular}
	 \caption{\label{tab:params}System parameters for simulation of $np_{3/2}$ states in Cs for two different single-photon excitations, S1 and S2, at temperature $T=300\;{\rm K}$. Lifetimes are calculated using expressions in \cite{Beterov_PRA_79_052504}. The relative blockades $b_{n,m}$ between Rydberg states $\ket{r}$ and $\ket{r_i}$ with principal quantum numbers $n,m$ are given in units of ${\sf B}_0$.}
\end{table}

\begin{figure}[!t]
 \centering
 \includegraphics[width=.95\linewidth]{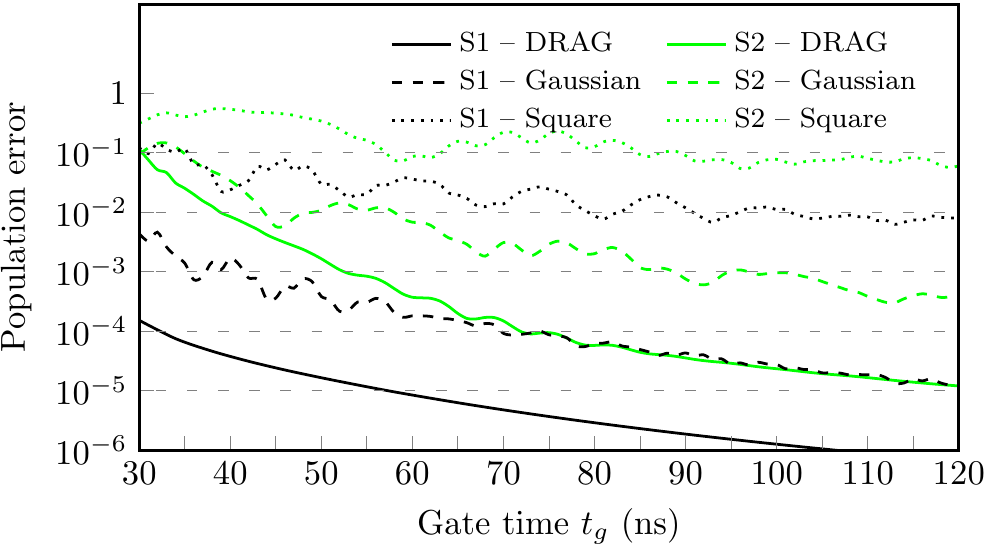}
 \caption{\label{fig:error_compare}(color online) Population error for a two-qubit Rydberg blockade entangling gate as a function of gate time $t_g=\tau_t+2\tau_c$. Gaussian pulses reduce leakage errors by up to $2.5$ orders of magnitude compared to conventional square controls, while additional supplementation with DRAG further improves by another $1.5$ orders of magnitude for reasonable gate times. The DRAG pulses are designed to minimize primarily leakage into the $\ket{n'}$ subset of the control atom as well as blockade leakage in the target atom. The Rydberg blockades ${\sf B}_0/2\pi$ are $1.54\;{\rm GHz}$ and $0.68\;{\rm GHz}$ for S1 and S2, respectively.}
\end{figure}

In \reffig{error_compare} we show the overall population error for a Rydberg blockade entangling gate according to the pulse sequence given in \reffig{sequence}a). Conventional square pulses perform very poorly due to a high degree of leakage. Gaussian pulses (we always compare Gaussian and DRAG pulses with equal values of $N$) show an improvement by $2$ to $2.5$ orders of magnitude over the square pulse sequence. This is attributed to Gaussians exponentially suppressing excitations to off-resonant transitions in the Fourier space, whilst square pulses only achieve a polynomial suppression.

Leakage can further be reduced by minimizing the main leakage channel into the $\ket{n'}$ subset of the control atom while also avoiding blockade leakage into the target Rydberg state $\ket{r}$ of the target atom with the aid of analytical DRAG pulses, whereby control and target pulses can be shaped independently of each other. Hence, for the area $\pi$ pulses we use $N=4$ in Eq.\refeq{ansatzDrag} to simultaneously suppress both $\del'$ transitions, whereas $N=2$ is sufficient for the $2\pi$ pulse since only leakage to the blockade-shifted target Rydberg state is significant. Note, however, that the error from the $2\pi$ pulse is more significant than that from the $\pi$ pulse since ${\sf B}_0$ is about half the value of $\del'$. Note also that we do not suppress the $\ket{n''}$ subset since this would require us to use $N=8$, which in turn increases the amplitude of the control pulse $\e_x$. The spectral argument that leads to our pulses only holds for $\e_x/\delta\ll 1$ ($\delta$ being the smallest detuning) so that increasing amplitudes deteriorate the gate. Owing to the frequencies $\del'$ and $\del''$ being very similar, the spectral power at both $\del''$ transitions is sufficiently low even though they are not explicitly nulled out, as can be seen in the spectrum shown in \reffig{sequence}d). Using these frequency-selective shapes additionally yields $1.5$ orders of magnitude improvement over Gaussians, hence improving over square pulses by up to four orders of magnitude. Best population errors are achieved for excitations in S1, owing to larger separations of atomic levels. Under these conditions, DRAG pulses allow speeding up gates by a factor of three compared to Gaussians, while achieving the same error. Compared to square pulses, the speed up lies in the range of several orders of magnitude.  

\subsection{Optimal Rydberg blockade} 
\begin{figure}
 \centering
 \includegraphics[width=.95\linewidth]{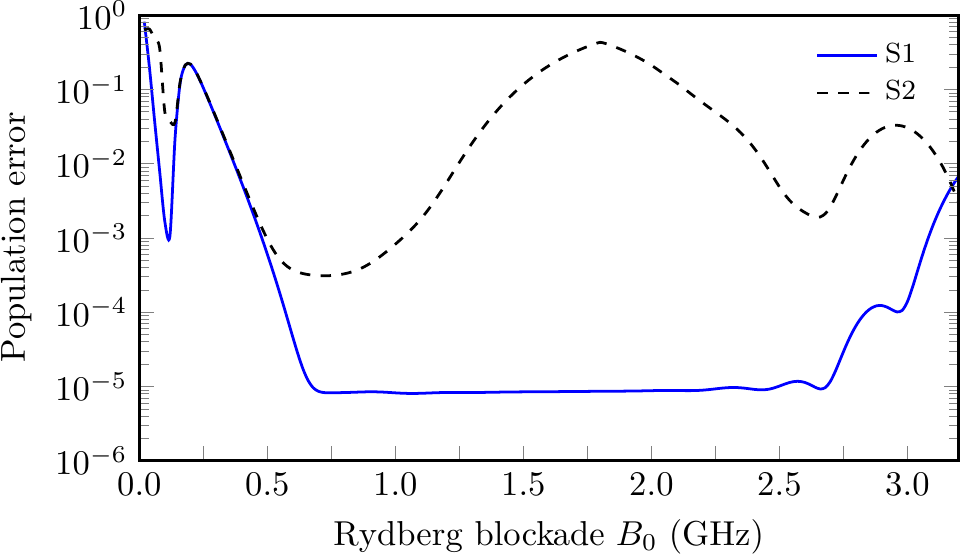}
 \caption{\label{fig:scan_B}(color online) Population error for a fixed gate time $\tau_t=30\;{\rm ns}$ as a function of the blockade shift ${\sf B}_0$ in settings S1 and S2. The error is minimized for a value of ${\sf B}_0/2\pi \sim 0.7\;{\rm GHz}$ in S2. In S1, the population error is optimal for blockade shifts in the range of $0.7-2.7\;{\rm GHz}$.}
\end{figure}

The performance of the Rydberg entangling gate strongly depends on the value of the blockade shifts. Scanning over the value of ${\sf B}_0$ for a fixed gate time ($\tau_t=30\;{\rm ns}$) reveals that the optimal value for ${\sf B}_0/2\pi$ is around $1.5(0.7)\;{\rm GHz}$, for settings S1(S2) as illustrated in \reffig{scan_B}. This is explained qualitatively by analyzing the energies of all involved Rydberg states. For this purpose, we assume for simplicity that all blockades ${\sf B}_{r_i,r_j}\sim {\sf B}_0$. Starting in the initial state $\rho_{in}=\ketbra{10}{10}$ we see that due to the first $\pi$-pulse populating $\ket{np_{3/2}}$, the Rydberg levels of the target atom are blockade-shifted by ${\sf B}_0$. As a consequence, for instance the leakage transitions into the $\ket{n''}$ subset are almost resonantly driven by the $2\pi$-pulse if ${\sf B}_0\sim (\del''_{1/2}+\del''_{3/2})/2$, leading to even more undesired excitation. On the other hand, too small a blockade will produce large population errors since the $2\pi$-pulse will leave  population inside the almost resonant blockade-shifted $\ket{np_{3/2}}$ state. This motivates a careful analysis of the Rydberg energies since an unsophisticated choice of ${\sf B}_0$ might introduce severe frequency crowding issues and tremendously lower gate fidelities.

Analytically estimating the optimal value for the blockade shift is possible by minimizing excitation to harmful levels\cite{Zhang_PRA_85_042310}. The matrix element for a transition to a Rydberg state $\ket{n}$ scales $\propto n^{-3/2}$. The probability of exciting states at detunings $\del_k$ scales $\propto \del_k^{-2}$ so that we can write the sum of all probabilities to excite harmful leakage states as
\begin{align}\label{eq:leak_prob}\begin{split}
	P_{\rm leak} & \propto \frac{1}{(n+1)^3(\del_1+{\sf B}_0)^2} + \frac{1}{(n-1)^3(\del_1-{\sf B}_0)^2}\\
	& + \frac{1}{(n')^3(\del_2-{\sf B}_0)^2} + \frac{1}{(n'')^3(\del_2+{\sf B}_0)^2}
	 + \frac{1}{n^3{\sf B}_0^2} .
\end{split}\end{align}
Here, we have set $\del_1 = (\del_++\del_-)/2$ and $\del_2 = (\del'_{1/2}+\del'_{3/2}+\del''_{1/2}+\del''_{3/2})/4$. Finding the roots of $\mathrm{d}P_{\rm leak}/\mathrm{d}{\sf B}_0$ in order to minimize Eq.\refeq{leak_prob} for e.g. setting S2, yields an optimal value for the blockade, ${\sf B}_0/2\pi \sim 0.68\;{\rm GHz}$ which is in very good agreement to the optimal value found numerically in \reffig{scan_B}. Note that the shape of $P_{\rm leak}$ may be very flat around its exact minimum. As a consequence, it may be possible for certain setups to achieve similar performance with blockades that are clearly below the analytical estimate, as we see from the blue line in \reffig{scan_B}. There, the analytical estimate is $1.54\;{\rm GHz}$ which is twice as much as the lowest optimal value found numerically. 

\subsection{Entanglement fidelity} 
\begin{figure}
 \centering
 \includegraphics[width=.95\linewidth]{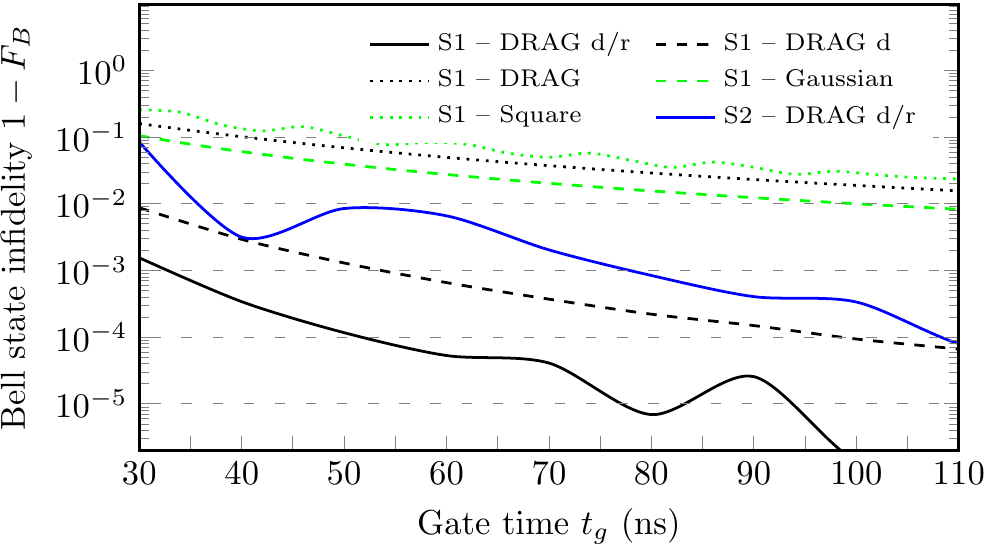}
 \caption{\label{fig:bellfid}(color online) Unitary Bell state infidelity as a measure for entanglement generated by the pulse sequence in \reffig{sequence} using square pulses, Gaussians, DRAG, detuned (d) DRAG controls and detuned DRAG controls with amplitude correction (d/r) for the setting S1 as well as optimized DRAG controls in S2. Detuning DRAG pulses on the target atom accounts for wrong phases and combines less leakage with high degrees of entanglement. The necessary detuning $\Lambda$ decreases proportionally to $1/\tau_t^2$ with a value of $\Lambda/2\pi=124.07\;{\rm MHz}$ at $\tau_t=25\;{\rm ns}$.}
\end{figure}

The ideal unitary after the sequence in \reffig{sequence} is
\begin{align}\label{eq:unit_cz_like} 
	\op{U}_{\mathrm{C_Z},\vec{\phi}} & = \mathrm{diag}(e^{i\phi_{00}},e^{i\phi_{01}},e^{i\phi_{10}},e^{i\phi_{11}}),
\end{align}
with $\phi_{ij}\equiv\phi_{ij,ij}$ being a shorthand notation for phases on the diagonal elements. To turn the $C_Z$-like gate in Eq.\refeq{unit_cz_like} into an entangling CNOT-like gate we slightly modify the procedure that turns a $C_Z$ into a CNOT gate: Applying a Hadamard on the target qubit before and after the $C_Z$ results in  a CNOT gate. Similarly, we find that a general $\pi/2$ rotation 
\begin{align}\label{eq:hadamardgen}
	\op{R}(\vec{h}) & = \frac{1}{\sqrt{2}}\begin{pmatrix}e^{ih_{00}} & e^{ih_{01}}\\e^{ih_{10}} & e^{ih_{11}}\end{pmatrix}
\end{align}
with phases $\vec{h}=(h_{00},h_{01},h_{10},h_{11})$ can be used to turn, up to relative phases, Eq.\refeq{unit_cz_like} into a CNOT. If the entangling phase $ \phi_{\rm ent}  = \phi_{00}-\phi_{01}-\phi_{10}+\phi_{11}$ of Eq.\refeq{unit_cz_like} is exactly $\pi$, the transformation
\begin{align}\label{eq:cnotlike}
	\left(\id\otimes\op{R}(\pi,\tilde{\phi},-\tilde{\phi},0)\right)\op{U}_{\mathrm{C_Z},\phi}\left(\id\otimes\op{R}(0,0,0,\pi)\right)
\end{align}
with $\tilde{\phi}=\phi_{10}-\phi_{11}$ produces a maximally entangling CNOT-like gate. In order to quantify the degree of entanglement of our pulse sequence, we pick $(\ket{00}+\ket{10})/\sqrt{2}$ as an initial state. Ideally, under Eq.\refeq{cnotlike} this yields, up to local phases, the maximally entangled Bell state $\ket{\Phi_+}=(\ket{00}+\ket{11})/\sqrt{2}$. To quantify the performance we evaluate the overlap fidelity between two density matrices $\rho,\rho_{\rm id}$ \cite{Raginsky_PLA_290_11}
\begin{align}\label{eq:fid_nu}
   F = \left(\PTrace{\mathbb{Q}}{\sqrt{\sqrt{\rho}\rho_{\rm id}\sqrt{\rho}}}\right)^2.
 \end{align}
Here, we take the partial trace over the computational subspace $\mathbb{Q}=\mathrm{span}\{\ket{00},\ket{01},\ket{10},\ket{11}\}$ to disregard irrelevant information about non-computational states. For $\rho_{\rm id}=\ketbra{\Phi_+}{\Phi_+}$ we denote the fidelity as Bell state fidelity $F_B$. The results are shown in the upper plot of \reffig{bellfid} whereby we assume that the $\pi/2$ gates on the qubit subspace are perfect gates. We observe that Gaussian controls seem to achieve better results than a naive DRAG control. However, the main reason for DRAG pulses to perform poorly at a first glance is wrong phases. Originally, it was proposed to change the drive frequency $\omega_d$ as a function of time to account for this effect \cite{Motzoi_PRL_103_110501}. However, it is also possible to employ a constant detuning $\Lambda$ from resonance, i.e. $\omega_d=\omega_r-\omega_q+\Lambda$ \cite{Theis_PRA_93_012324}, with the benefit of less experimental effort being required. We find, that detuning the target $2\pi$ pulse is sufficient to achieve low enough errors. As a consequence of off-resonant drive, rotation errors will be induced which can be corrected by rescaling the amplitudes of the pulses (by up to $3\%$ only for the fastest gates). The difference between the solid black line and the dotted red one in \reffig{bellfid} illustrates that a constant detuning and a rescaled amplitude indeed account for this induced error, yielding at least two orders of magnitude improvement over Gaussian waveforms. As one would expect from previous results \cite{Motzoi_PRL_103_110501}, the detuning scales proportionally to the Rabi frequency squared, yielding approximately a $1/\tau_t^2$ power law whereby the optimal detuning for a $2\pi$ pulse of $25\;{\rm ns}$ is $124.07\;{\rm MHz}$. We find that we are able to produce Bell states with a fidelity of $0.9999$ for $t_g\sim 50\;{\rm ns}$ using detuned DRAG pulses with amplitude correction. 

An alternate approach to account for phase issues is by waiting an appropriate time $t_{\rm gap}$ between the pulses \cite{Maller_PRA_92_022336} or to track phases in software and correct for them afterwards. The former approach will noticeably prolong the gate times compared to our approach. Overall, detuned DRAG pulses yield an improvement of more than two orders of magnitude compared to conventional shapes. Furthermore, the necessary gate times are less than $10^{-7}$ of the few second coherence times that have been demonstrated with neutral atom qubits\cite{YWang2015}, substantiating that Rydberg gates  are a promising approach for scalable quantum computing.

\subsection{Including spontaneous emission}  
\begin{figure}
 \centering
 \includegraphics[width=.95\linewidth]{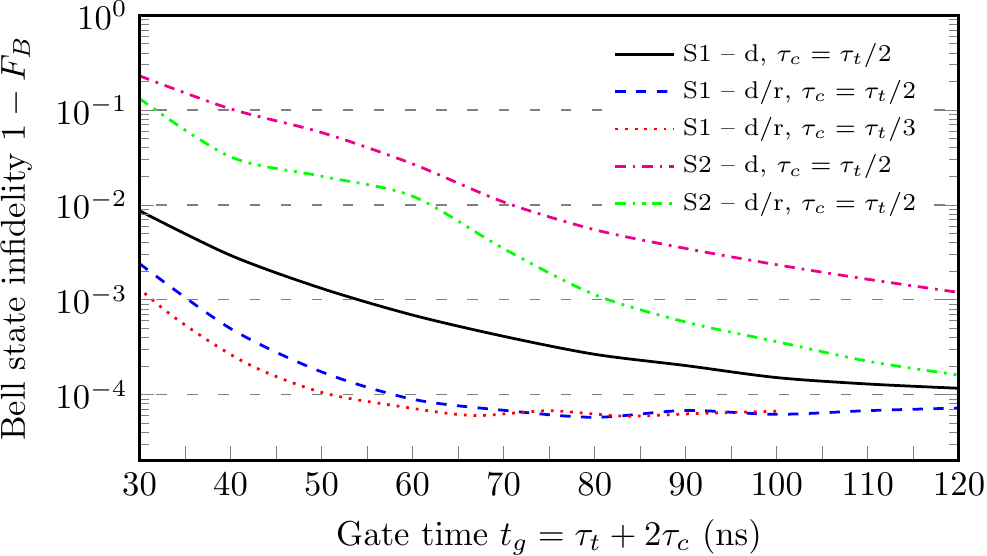}
 \caption{\label{fig:bellfid_nu}(color online) Bell state infidelity including decay from all Rydberg levels for optimized detuned DRAG controls in both settings S1 and S2. Bell states are generated with a fidelity of $0.9999$ at a total gate time of only $50\;{\rm ns}$.}
\end{figure}

All results in the previous section are based on unitary evolution of the atoms. A more realistic model incorporates decay due to finite lifetimes of the energy levels. We employ a Lindbladian model to simulate the effects of  decoherence, whereby we assume that population of Rydberg levels decays by a fraction of $7/8$ into some auxiliary level $\ket{g}$ that has zero effect on the rest of the dynamics. The residual part decays with equal probabilities into the states $\ket{0}$ and $\ket{1}$ of the atoms. Hence, the full dynamics of our system are goverened by the Lindblad master equation for the density operator $\op{\rho}$
\begin{align}
  \dot{\op{\rho}} & = -i\comut{\op{H}}{\op{\rho}} -\frac{1}{2}\sum\limits_r \left(\op{C}_r^{\dagger}\op{C}_r\rho + \rho\op{C}_r^{\dagger}\op{C}\right) + \sum\limits_r\op{C}_r\op{\rho}\op{C}_r^{\dagger}.
\end{align}
Here, the operators $\op{C}_r=\op{c}_r \otimes \id + \id \otimes \op{c}_r$ describe decay of all relevant Rydberg states $\ket{r}$ in both atoms into $\ket{g}$, $\ket{0}$ and $\ket{1}$, i.e.
\begin{align}
 \op{c}_r & = \sqrt{\Gamma_r}\LR{\frac{7}{8}\ketbra{g}{r}+\frac{1}{16}\ketbra{0}{r}+\frac{1}{16}\ketbra{1}{r}}.
\end{align}
The decay rate $\Gamma_r$ is the inverse of the lifetime $\tau_r$ of a Rydberg state $\ket{r}$. Values for the target Rydberg states in both settings are given in \reftab{params}. For an experiment at room temperature ($\sim 300\;{\rm K}$) in setting S1, we find that Bell states are generated with a fidelity of better than $0.9999$ at a gate time of $\lesssim 60\;{\rm ns}$. The results for optimized DRAG pulses are plotted in \reffig{bellfid_nu}. As expected because of shorter lifetimes, non-unitary errors become visible earlier in S1 than in S2. However, unitary errors are dominant, so that gates in S1 apear to be more promising than those in S2, despite the shorter lifetimes. Since the $\pi$ pulses on the control atom do not require blockade effects, we may run them faster without losing performance. The dotted red curve in \reffig{bellfid_nu} confirms this observation. For $\tau_c=\tau_t/3$ we achieve slightly better results, yielding errors less than $10^{-4}$ at only $50\;{\rm ns}$ gate time. In a $4\;{\rm K}$ environment lifetimes will be on the order of a few ms, allowing for performance very similar to that for the unitary analysis. 

We have characterized the gate performance in terms of the Bell state fidelity. While the fidelity is the most widely used measure of gate performance, others   have been proposed\cite{Gilchrist2005}. In particular the trace distance 
has been shown to be linearly sensitive to Rydberg gate phase errors that affect the fidelity only quadratically\cite{Zhang_PRA_85_042310}.
Using the rescaled and detuned DRAG gates that optimize the fidelity we find that the trace distance error is an order of magnitude larger. 
As it is an open question as to which performance measure is most relevant for specific quantum computational tasks  we have not studied the trace distance in more detail, although  we anticipate that the trace distance error could also be reduced with appropriate pulse design, 

\section{Summary} In conclusion we have presented DRAG pulses with $x$ quadrature  control for Rydberg blockade gates that lead to Bell state fidelity $F_B>0.9999$ with gate times of 50 ns. The pulses are generated with an analytical method that could readily be extended to the level structure of other atoms.  The results fully account for all the dominant leakage channels as well as Rydberg decay in a room temperature environment.  The 50 ns gate time is orders of magnitude faster than high fidelity trapped ion gates, about the same speed as state of the art superconducting qubit gates, while the ratio of coherence time to gate time is orders of magnitude better.
Together with recent progress in high fidelity single qubit gates\cite{Xia2015,YWang2015} DRAG pulses establish neutral atom qubits with Rydberg gates as a promising candidate for scalable quantum computation. Our result specifically applies to the case of one-photon Rydberg excitation. We leave extension to the more common case of two-photon excitation for future work. We also emphasize that the predicted gate fidelity assumes no technical errors and ground state laser cooling. Demonstrating real performance close to the theoretical level established here remains an outstanding challenge. 

\section*{Acknowledgments}
MS acknowledges funding from the IARPA MQCO program through ARO contract W911NF-10-1-0347, the ARL-CDQI through cooperative agreement W911NF-15-2-0061 and NSF  award 1521374. LST, FM and FKW were supported by the European Union through SCALEQIT.

%
\bibliography{Bibliography.bib}

\begin{thebibliography}{29}%
\makeatletter
\providecommand \@ifxundefined [1]{%
 \@ifx{#1\undefined}
}%
\providecommand \@ifnum [1]{%
 \ifnum #1\expandafter \@firstoftwo
 \else \expandafter \@secondoftwo
 \fi
}%
\providecommand \@ifx [1]{%
 \ifx #1\expandafter \@firstoftwo
 \else \expandafter \@secondoftwo
 \fi
}%
\providecommand \natexlab [1]{#1}%
\providecommand \enquote  [1]{``#1''}%
\providecommand \bibnamefont  [1]{#1}%
\providecommand \bibfnamefont [1]{#1}%
\providecommand \citenamefont [1]{#1}%
\providecommand \href@noop [0]{\@secondoftwo}%
\providecommand \href [0]{\begingroup \@sanitize@url \@href}%
\providecommand \@href[1]{\@@startlink{#1}\@@href}%
\providecommand \@@href[1]{\endgroup#1\@@endlink}%
\providecommand \@sanitize@url [0]{\catcode `\\12\catcode `\$12\catcode
  `\&12\catcode `\#12\catcode `\^12\catcode `\_12\catcode `\%12\relax}%
\providecommand \@@startlink[1]{}%
\providecommand \@@endlink[0]{}%
\providecommand \url  [0]{\begingroup\@sanitize@url \@url }%
\providecommand \@url [1]{\endgroup\@href {#1}{\urlprefix }}%
\providecommand \urlprefix  [0]{URL }%
\providecommand \Eprint [0]{\href }%
\providecommand \doibase [0]{http://dx.doi.org/}%
\providecommand \selectlanguage [0]{\@gobble}%
\providecommand \bibinfo  [0]{\@secondoftwo}%
\providecommand \bibfield  [0]{\@secondoftwo}%
\providecommand \translation [1]{[#1]}%
\providecommand \BibitemOpen [0]{}%
\providecommand \bibitemStop [0]{}%
\providecommand \bibitemNoStop [0]{.\EOS\space}%
\providecommand \EOS [0]{\spacefactor3000\relax}%
\providecommand \BibitemShut  [1]{\csname bibitem#1\endcsname}%
\let\auto@bib@innerbib\@empty
\bibitem [{\citenamefont {Jaksch}\ \emph {et~al.}(2000)\citenamefont {Jaksch},
  \citenamefont {Cirac}, \citenamefont {Zoller}, \citenamefont {Rolston},
  \citenamefont {C\^ot\'e},\ and\ \citenamefont {Lukin}}]{Jaksch2000}%
  \BibitemOpen
  \bibfield  {author} {\bibinfo {author} {\bibfnamefont {D.}~\bibnamefont
  {Jaksch}}, \bibinfo {author} {\bibfnamefont {J.~I.}\ \bibnamefont {Cirac}},
  \bibinfo {author} {\bibfnamefont {P.}~\bibnamefont {Zoller}}, \bibinfo
  {author} {\bibfnamefont {S.~L.}\ \bibnamefont {Rolston}}, \bibinfo {author}
  {\bibfnamefont {R.}~\bibnamefont {C\^ot\'e}}, \ and\ \bibinfo {author}
  {\bibfnamefont {M.~D.}\ \bibnamefont {Lukin}},\ }\href
  {http://dx.doi.org/10.1103/PhysRevLett.85.2208} {\bibfield  {journal}
  {\bibinfo  {journal} {Phys. Rev. Lett.}\ }\textbf {\bibinfo {volume} {85}},\
  \bibinfo {pages} {2208} (\bibinfo {year} {2000})}\BibitemShut {NoStop}%
\bibitem [{\citenamefont {Maller}\ \emph {et~al.}(2015)\citenamefont {Maller},
  \citenamefont {Lichtman}, \citenamefont {Xia}, \citenamefont {Sun},
  \citenamefont {Piotrowicz}, \citenamefont {Carr}, \citenamefont {Isenhower},\
  and\ \citenamefont {Saffman}}]{Maller_PRA_92_022336}%
  \BibitemOpen
  \bibfield  {author} {\bibinfo {author} {\bibfnamefont {K.~M.}\ \bibnamefont
  {Maller}}, \bibinfo {author} {\bibfnamefont {M.~T.}\ \bibnamefont
  {Lichtman}}, \bibinfo {author} {\bibfnamefont {T.}~\bibnamefont {Xia}},
  \bibinfo {author} {\bibfnamefont {Y.}~\bibnamefont {Sun}}, \bibinfo {author}
  {\bibfnamefont {M.~J.}\ \bibnamefont {Piotrowicz}}, \bibinfo {author}
  {\bibfnamefont {A.~W.}\ \bibnamefont {Carr}}, \bibinfo {author}
  {\bibfnamefont {L.}~\bibnamefont {Isenhower}}, \ and\ \bibinfo {author}
  {\bibfnamefont {M.}~\bibnamefont {Saffman}},\ }\href
  {http://dx.doi.org/10.1103/PhysRevA.92.022336} {\bibfield  {journal}
  {\bibinfo  {journal} {Phys. Rev. A}\ }\textbf {\bibinfo {volume} {92}},\
  \bibinfo {pages} {022336} (\bibinfo {year} {2015})}\BibitemShut {NoStop}%
\bibitem [{\citenamefont {Jau}\ \emph {et~al.}(2016)\citenamefont {Jau},
  \citenamefont {Hankin}, \citenamefont {Keating}, \citenamefont {Deutsch},\
  and\ \citenamefont {Biedermann}}]{Jau2016}%
  \BibitemOpen
  \bibfield  {author} {\bibinfo {author} {\bibfnamefont {Y.-Y.}\ \bibnamefont
  {Jau}}, \bibinfo {author} {\bibfnamefont {A.~M.}\ \bibnamefont {Hankin}},
  \bibinfo {author} {\bibfnamefont {T.}~\bibnamefont {Keating}}, \bibinfo
  {author} {\bibfnamefont {I.~H.}\ \bibnamefont {Deutsch}}, \ and\ \bibinfo
  {author} {\bibfnamefont {G.~W.}\ \bibnamefont {Biedermann}},\ }\href
  {http://dx.doi.org/10.1038/nphys3487} {\bibfield  {journal} {\bibinfo
  {journal} {Nat. Phys.}\ }\textbf {\bibinfo {volume} {12}},\ \bibinfo {pages}
  {71} (\bibinfo {year} {2016})}\BibitemShut {NoStop}%
\bibitem [{\citenamefont {Saffman}(2016)}]{Saffman2016}%
  \BibitemOpen
  \bibfield  {author} {\bibinfo {author} {\bibfnamefont {M.}~\bibnamefont
  {Saffman}},\ }\href {https://arxiv.org/abs/1605.05207} {\bibfield  {journal}
  {\bibinfo  {journal} {arXiv:1605.05207}\ } (\bibinfo {year}
  {2016})}\BibitemShut {NoStop}%
\bibitem [{\citenamefont {Zhang}\ \emph {et~al.}(2012)\citenamefont {Zhang},
  \citenamefont {Gill}, \citenamefont {Isenhower}, \citenamefont {Walker},\
  and\ \citenamefont {Saffman}}]{Zhang_PRA_85_042310}%
  \BibitemOpen
  \bibfield  {author} {\bibinfo {author} {\bibfnamefont {X.~L.}\ \bibnamefont
  {Zhang}}, \bibinfo {author} {\bibfnamefont {A.~T.}\ \bibnamefont {Gill}},
  \bibinfo {author} {\bibfnamefont {L.}~\bibnamefont {Isenhower}}, \bibinfo
  {author} {\bibfnamefont {T.~G.}\ \bibnamefont {Walker}}, \ and\ \bibinfo
  {author} {\bibfnamefont {M.}~\bibnamefont {Saffman}},\ }\href
  {http://dx.doi.org/10.1103/PhysRevA.85.042310} {\bibfield  {journal}
  {\bibinfo  {journal} {Phys. Rev. A}\ }\textbf {\bibinfo {volume} {85}},\
  \bibinfo {pages} {042310} (\bibinfo {year} {2012})}\BibitemShut {NoStop}%
\bibitem [{\citenamefont {M\"uller}\ \emph {et~al.}(2011)\citenamefont
  {M\"uller}, \citenamefont {Haakh}, \citenamefont {Calarco}, \citenamefont
  {Koch},\ and\ \citenamefont {Henkel}}]{Muller2011}%
  \BibitemOpen
  \bibfield  {author} {\bibinfo {author} {\bibfnamefont {M.~M.}\ \bibnamefont
  {M\"uller}}, \bibinfo {author} {\bibfnamefont {H.~R.}\ \bibnamefont {Haakh}},
  \bibinfo {author} {\bibfnamefont {T.}~\bibnamefont {Calarco}}, \bibinfo
  {author} {\bibfnamefont {C.~P.}\ \bibnamefont {Koch}}, \ and\ \bibinfo
  {author} {\bibfnamefont {C.}~\bibnamefont {Henkel}},\ }\href
  {http://dx.doi.org/10.1007/s11128-011-0296-0} {\bibfield  {journal} {\bibinfo
   {journal} {Quant. Inf. Proc.}\ }\textbf {\bibinfo {volume} {10}},\ \bibinfo
  {pages} {771} (\bibinfo {year} {2011})}\BibitemShut {NoStop}%
\bibitem [{\citenamefont {Goerz}\ \emph {et~al.}(2014)\citenamefont {Goerz},
  \citenamefont {Halperin}, \citenamefont {Aytac}, \citenamefont {Koch},\ and\
  \citenamefont {Whaley}}]{Goerz_PRA_90_032329}%
  \BibitemOpen
  \bibfield  {author} {\bibinfo {author} {\bibfnamefont {M.~H.}\ \bibnamefont
  {Goerz}}, \bibinfo {author} {\bibfnamefont {E.~J.}\ \bibnamefont {Halperin}},
  \bibinfo {author} {\bibfnamefont {J.~M.}\ \bibnamefont {Aytac}}, \bibinfo
  {author} {\bibfnamefont {C.~P.}\ \bibnamefont {Koch}}, \ and\ \bibinfo
  {author} {\bibfnamefont {K.~B.}\ \bibnamefont {Whaley}},\ }\href
  {http://link.aps.org/doi/10.1103/PhysRevA.90.032329} {\bibfield  {journal}
  {\bibinfo  {journal} {Phys. Rev. A}\ }\textbf {\bibinfo {volume} {90}},\
  \bibinfo {pages} {032329} (\bibinfo {year} {2014})}\BibitemShut {NoStop}%
\bibitem [{\citenamefont {M\"uller}\ \emph {et~al.}(2014)\citenamefont
  {M\"uller}, \citenamefont {Murphy}, \citenamefont {Montangero}, \citenamefont
  {Calarco}, \citenamefont {Grangier},\ and\ \citenamefont
  {Browaeys}}]{Muller_PRA_89_032334}%
  \BibitemOpen
  \bibfield  {author} {\bibinfo {author} {\bibfnamefont {M.~M.}\ \bibnamefont
  {M\"uller}}, \bibinfo {author} {\bibfnamefont {M.}~\bibnamefont {Murphy}},
  \bibinfo {author} {\bibfnamefont {S.}~\bibnamefont {Montangero}}, \bibinfo
  {author} {\bibfnamefont {T.}~\bibnamefont {Calarco}}, \bibinfo {author}
  {\bibfnamefont {P.}~\bibnamefont {Grangier}}, \ and\ \bibinfo {author}
  {\bibfnamefont {A.}~\bibnamefont {Browaeys}},\ }\href
  {http://link.aps.org/doi/10.1103/PhysRevA.89.032334} {\bibfield  {journal}
  {\bibinfo  {journal} {Phys. Rev. A}\ }\textbf {\bibinfo {volume} {89}},\
  \bibinfo {pages} {032334} (\bibinfo {year} {2014})}\BibitemShut {NoStop}%
\bibitem [{\citenamefont {Rao}\ and\ \citenamefont
  {M\o{}lmer}(2014)}]{Rao2014}%
  \BibitemOpen
  \bibfield  {author} {\bibinfo {author} {\bibfnamefont {D.~D.~B.}\
  \bibnamefont {Rao}}\ and\ \bibinfo {author} {\bibfnamefont {K.}~\bibnamefont
  {M\o{}lmer}},\ }\href {http://dx.doi.org/10.1103/PhysRevA.89.030301}
  {\bibfield  {journal} {\bibinfo  {journal} {Phys. Rev. A}\ }\textbf {\bibinfo
  {volume} {89}},\ \bibinfo {pages} {030301(R)} (\bibinfo {year}
  {2014})}\BibitemShut {NoStop}%
\bibitem [{\citenamefont {Han}\ \emph {et~al.}(2016)\citenamefont {Han},
  \citenamefont {Ng},\ and\ \citenamefont {Englert}}]{Han2016}%
  \BibitemOpen
  \bibfield  {author} {\bibinfo {author} {\bibfnamefont {R.}~\bibnamefont
  {Han}}, \bibinfo {author} {\bibfnamefont {H.~K.}\ \bibnamefont {Ng}}, \ and\
  \bibinfo {author} {\bibfnamefont {B.-G.}\ \bibnamefont {Englert}},\ }\href
  {http://dx.doi.org/10.1209/0295-5075/113/40001} {\bibfield  {journal}
  {\bibinfo  {journal} {Europhys. Lett.}\ }\textbf {\bibinfo {volume} {113}},\
  \bibinfo {pages} {40001} (\bibinfo {year} {2016})}\BibitemShut {NoStop}%
\bibitem [{\citenamefont {Su}\ \emph {et~al.}(2016)\citenamefont {Su},
  \citenamefont {Liang}, \citenamefont {Zhang}, \citenamefont {Wen},
  \citenamefont {Sun}, \citenamefont {Jin},\ and\ \citenamefont
  {Zhu}}]{Su2016}%
  \BibitemOpen
  \bibfield  {author} {\bibinfo {author} {\bibfnamefont {S.-L.}\ \bibnamefont
  {Su}}, \bibinfo {author} {\bibfnamefont {E.}~\bibnamefont {Liang}}, \bibinfo
  {author} {\bibfnamefont {S.}~\bibnamefont {Zhang}}, \bibinfo {author}
  {\bibfnamefont {J.-J.}\ \bibnamefont {Wen}}, \bibinfo {author} {\bibfnamefont
  {L.-L.}\ \bibnamefont {Sun}}, \bibinfo {author} {\bibfnamefont
  {Z.}~\bibnamefont {Jin}}, \ and\ \bibinfo {author} {\bibfnamefont {A.-D.}\
  \bibnamefont {Zhu}},\ }\href {http://dx.doi.org/10.1103/PhysRevA.93.012306}
  {\bibfield  {journal} {\bibinfo  {journal} {Phys. Rev. A}\ }\textbf {\bibinfo
  {volume} {93}},\ \bibinfo {pages} {012306} (\bibinfo {year}
  {2016})}\BibitemShut {NoStop}%
\bibitem [{\citenamefont {Devitt}\ \emph {et~al.}(2013)\citenamefont {Devitt},
  \citenamefont {Munro},\ and\ \citenamefont {Nemoto}}]{Devitt2013}%
  \BibitemOpen
  \bibfield  {author} {\bibinfo {author} {\bibfnamefont {S.~J.}\ \bibnamefont
  {Devitt}}, \bibinfo {author} {\bibfnamefont {W.~J.}\ \bibnamefont {Munro}}, \
  and\ \bibinfo {author} {\bibfnamefont {K.}~\bibnamefont {Nemoto}},\ }\href
  {http://iopscience.iop.org/article/10.1088/0034-4885/76/7/076001/meta}
  {\bibfield  {journal} {\bibinfo  {journal} {Rep. Prog. Phys.}\ }\textbf
  {\bibinfo {volume} {76}},\ \bibinfo {pages} {076001} (\bibinfo {year}
  {2013})}\BibitemShut {NoStop}%
\bibitem [{\citenamefont {Motzoi}\ \emph {et~al.}(2009)\citenamefont {Motzoi},
  \citenamefont {Gambetta}, \citenamefont {Rebentrost},\ and\ \citenamefont
  {Wilhelm}}]{Motzoi_PRL_103_110501}%
  \BibitemOpen
  \bibfield  {author} {\bibinfo {author} {\bibfnamefont {F.}~\bibnamefont
  {Motzoi}}, \bibinfo {author} {\bibfnamefont {J.~M.}\ \bibnamefont
  {Gambetta}}, \bibinfo {author} {\bibfnamefont {P.}~\bibnamefont
  {Rebentrost}}, \ and\ \bibinfo {author} {\bibfnamefont {F.~K.}\ \bibnamefont
  {Wilhelm}},\ }\href {http://dx.doi.org/10.1103/PhysRevLett.103.110501}
  {\bibfield  {journal} {\bibinfo  {journal} {Phys. Rev. Lett.}\ }\textbf
  {\bibinfo {volume} {103}},\ \bibinfo {pages} {110501} (\bibinfo {year}
  {2009})}\BibitemShut {NoStop}%
\bibitem [{\citenamefont {Egger}\ and\ \citenamefont
  {Wilhelm}(2013)}]{Egger_SUST_2013}%
  \BibitemOpen
  \bibfield  {author} {\bibinfo {author} {\bibfnamefont {D.~J.}\ \bibnamefont
  {Egger}}\ and\ \bibinfo {author} {\bibfnamefont {F.~K.}\ \bibnamefont
  {Wilhelm}},\ }\href {http://dx.doi.org/10.1088/0953-2048/27/1/014001}
  {\bibfield  {journal} {\bibinfo  {journal} {Supercond. Sci. Technol.}\
  }\textbf {\bibinfo {volume} {27}},\ \bibinfo {pages} {014001} (\bibinfo
  {year} {2013})}\BibitemShut {NoStop}%
\bibitem [{\citenamefont {Ghosh}\ \emph {et~al.}(2013)\citenamefont {Ghosh},
  \citenamefont {Galiautdinov}, \citenamefont {Zhou}, \citenamefont {Korotkov},
  \citenamefont {Martinis},\ and\ \citenamefont
  {Geller}}]{Ghosh_PRA_87_022309}%
  \BibitemOpen
  \bibfield  {author} {\bibinfo {author} {\bibfnamefont {J.}~\bibnamefont
  {Ghosh}}, \bibinfo {author} {\bibfnamefont {A.}~\bibnamefont {Galiautdinov}},
  \bibinfo {author} {\bibfnamefont {Z.}~\bibnamefont {Zhou}}, \bibinfo {author}
  {\bibfnamefont {A.~N.}\ \bibnamefont {Korotkov}}, \bibinfo {author}
  {\bibfnamefont {J.~M.}\ \bibnamefont {Martinis}}, \ and\ \bibinfo {author}
  {\bibfnamefont {M.~R.}\ \bibnamefont {Geller}},\ }\href
  {http://dx.doi.org/10.1103/PhysRevA.87.022309} {\bibfield  {journal}
  {\bibinfo  {journal} {Phys. Rev. A}\ }\textbf {\bibinfo {volume} {87}},\
  \bibinfo {pages} {022309} (\bibinfo {year} {2013})}\BibitemShut {NoStop}%
\bibitem [{\citenamefont {Fermi}(1930)}]{Fermi1930}%
  \BibitemOpen
  \bibfield  {author} {\bibinfo {author} {\bibfnamefont {E.}~\bibnamefont
  {Fermi}},\ }\href {http://link.springer.com/article/10.1007%2FBF01344810}
  {\bibfield  {journal} {\bibinfo  {journal} {Z. Phys.}\ }\textbf {\bibinfo
  {volume} {59}},\ \bibinfo {pages} {680} (\bibinfo {year} {1930})}\BibitemShut
  {NoStop}%
\bibitem [{\citenamefont {Lorenzen}\ and\ \citenamefont
  {Niemax}(1978)}]{Lorenzen1978}%
  \BibitemOpen
  \bibfield  {author} {\bibinfo {author} {\bibfnamefont {C.~J.}\ \bibnamefont
  {Lorenzen}}\ and\ \bibinfo {author} {\bibfnamefont {K.}~\bibnamefont
  {Niemax}},\ }\href {http://dx.doi.org/10.1088/0022-3700/11/23/003} {\bibfield
   {journal} {\bibinfo  {journal} {J. Phys. B}\ }\textbf {\bibinfo {volume}
  {11}},\ \bibinfo {pages} {L723} (\bibinfo {year} {1978})}\BibitemShut
  {NoStop}%
\bibitem [{\citenamefont {Motzoi}\ and\ \citenamefont
  {Wilhelm}(2013)}]{Motzoi_PRA_88_062318}%
  \BibitemOpen
  \bibfield  {author} {\bibinfo {author} {\bibfnamefont {F.}~\bibnamefont
  {Motzoi}}\ and\ \bibinfo {author} {\bibfnamefont {F.~K.}\ \bibnamefont
  {Wilhelm}},\ }\href {http://dx.doi.org/10.1103/PhysRevA.88.062318} {\bibfield
   {journal} {\bibinfo  {journal} {Phys. Rev. A}\ }\textbf {\bibinfo {volume}
  {88}},\ \bibinfo {pages} {062318} (\bibinfo {year} {2013})}\BibitemShut
  {NoStop}%
\bibitem [{\citenamefont {Demirplak}\ and\ \citenamefont
  {Rice}(2003)}]{Demirplak_Chem_107_9937}%
  \BibitemOpen
  \bibfield  {author} {\bibinfo {author} {\bibfnamefont {M.}~\bibnamefont
  {Demirplak}}\ and\ \bibinfo {author} {\bibfnamefont {S.~A.}\ \bibnamefont
  {Rice}},\ }\href {\doibase http://dx.doi.org/10.1021/jp030708a} {\bibfield
  {journal} {\bibinfo  {journal} {J. Phys. Chem. A}\ }\textbf {\bibinfo
  {volume} {107}},\ \bibinfo {pages} {9937} (\bibinfo {year}
  {2003})}\BibitemShut {NoStop}%
\bibitem [{\citenamefont {Torrontegui}\ \emph {et~al.}(2013)\citenamefont
  {Torrontegui}, \citenamefont {Ib{\'a}nez}, \citenamefont
  {Mart{\'\i}nez-Garaot}, \citenamefont {Modugno}, \citenamefont {del Campo},
  \citenamefont {Gu{\'e}ry-Odelin}, \citenamefont {Ruschhaupt}, \citenamefont
  {Chen}, \citenamefont {Muga} \emph {et~al.}}]{Torrontegui_shortcuts}%
  \BibitemOpen
  \bibfield  {author} {\bibinfo {author} {\bibfnamefont {E.}~\bibnamefont
  {Torrontegui}}, \bibinfo {author} {\bibfnamefont {S.}~\bibnamefont
  {Ib{\'a}nez}}, \bibinfo {author} {\bibfnamefont {S.}~\bibnamefont
  {Mart{\'\i}nez-Garaot}}, \bibinfo {author} {\bibfnamefont {M.}~\bibnamefont
  {Modugno}}, \bibinfo {author} {\bibfnamefont {A.}~\bibnamefont {del Campo}},
  \bibinfo {author} {\bibfnamefont {D.}~\bibnamefont {Gu{\'e}ry-Odelin}},
  \bibinfo {author} {\bibfnamefont {A.}~\bibnamefont {Ruschhaupt}}, \bibinfo
  {author} {\bibfnamefont {X.}~\bibnamefont {Chen}}, \bibinfo {author}
  {\bibfnamefont {J.~G.}\ \bibnamefont {Muga}},  \emph {et~al.},\ }\href
  {\doibase http://dx.doi.org/10.1016/B978-0-12-408090-4.00002-5} {\bibfield
  {journal} {\bibinfo  {journal} {Adv. At. Mol. Opt. Phys}\ }\textbf {\bibinfo
  {volume} {62}},\ \bibinfo {pages} {117} (\bibinfo {year} {2013})}\BibitemShut
  {NoStop}%
\bibitem [{\citenamefont {Warren}(1984)}]{Warren_JChemPhys_81_5437}%
  \BibitemOpen
  \bibfield  {author} {\bibinfo {author} {\bibfnamefont {W.~S.}\ \bibnamefont
  {Warren}},\ }\href {http://dx.doi.org/10.1063/1.447644} {\bibfield  {journal}
  {\bibinfo  {journal} {J. Chem. Phys}\ }\textbf {\bibinfo {volume} {81}},\
  \bibinfo {pages} {5437} (\bibinfo {year} {1984})}\BibitemShut {NoStop}%
\bibitem [{\citenamefont {Gambetta}\ \emph {et~al.}(2011)\citenamefont
  {Gambetta}, \citenamefont {Motzoi}, \citenamefont {Merkel},\ and\
  \citenamefont {Wilhelm}}]{Gambetta_PRA_83_012308}%
  \BibitemOpen
  \bibfield  {author} {\bibinfo {author} {\bibfnamefont {J.~M.}\ \bibnamefont
  {Gambetta}}, \bibinfo {author} {\bibfnamefont {F.}~\bibnamefont {Motzoi}},
  \bibinfo {author} {\bibfnamefont {S.~T.}\ \bibnamefont {Merkel}}, \ and\
  \bibinfo {author} {\bibfnamefont {F.~K.}\ \bibnamefont {Wilhelm}},\ }\href
  {http://dx.doi.org/10.1103/PhysRevA.83.012308} {\bibfield  {journal}
  {\bibinfo  {journal} {Phys. Rev. A}\ }\textbf {\bibinfo {volume} {83}},\
  \bibinfo {pages} {012308} (\bibinfo {year} {2011})}\BibitemShut {NoStop}%
\bibitem [{alp()}]{alphanote}%
  \BibitemOpen
  \href@noop {} {}\bibinfo {note} {The AC polarizabilities of the $6s_{1/2}$
  ground state and Rydberg states are slightly different which leads to a time
  dependent detuning during the DRAG pulse. This has not been included in the
  simulations but can be fully corrected for by changing $\omega_d$ during the
  pulse or by slightly changing the detuning $\Lambda$ introduced after
  Eq.\refeq{fid_nu}.}\BibitemShut {Stop}%
\bibitem [{\citenamefont {Beterov}\ \emph {et~al.}(2009)\citenamefont
  {Beterov}, \citenamefont {Ryabtsev}, \citenamefont {Tretyakov},\ and\
  \citenamefont {Entin}}]{Beterov_PRA_79_052504}%
  \BibitemOpen
  \bibfield  {author} {\bibinfo {author} {\bibfnamefont {I.~I.}\ \bibnamefont
  {Beterov}}, \bibinfo {author} {\bibfnamefont {I.~I.}\ \bibnamefont
  {Ryabtsev}}, \bibinfo {author} {\bibfnamefont {D.~B.}\ \bibnamefont
  {Tretyakov}}, \ and\ \bibinfo {author} {\bibfnamefont {V.~M.}\ \bibnamefont
  {Entin}},\ }\href {http://dx.doi.org/10.1103/PhysRevA.79.052504} {\bibfield
  {journal} {\bibinfo  {journal} {Phys. Rev. A}\ }\textbf {\bibinfo {volume}
  {79}},\ \bibinfo {pages} {052504} (\bibinfo {year} {2009})}\BibitemShut
  {NoStop}%
\bibitem [{\citenamefont {Raginsky}(2001)}]{Raginsky_PLA_290_11}%
  \BibitemOpen
  \bibfield  {author} {\bibinfo {author} {\bibfnamefont {M.}~\bibnamefont
  {Raginsky}},\ }\href
  {http://arxiv.org/ct?url=http%3A%2F%2Fdx.doi.org%2F10%252E1016%2FS0375-9601%252801%252900640-5&v=728200b7}
  {\bibfield  {journal} {\bibinfo  {journal} {Physics Letters A}\ }\textbf
  {\bibinfo {volume} {290}},\ \bibinfo {pages} {11} (\bibinfo {year}
  {2001})}\BibitemShut {NoStop}%
\bibitem [{\citenamefont {Theis}\ \emph {et~al.}(2016)\citenamefont {Theis},
  \citenamefont {Motzoi},\ and\ \citenamefont {Wilhelm}}]{Theis_PRA_93_012324}%
  \BibitemOpen
  \bibfield  {author} {\bibinfo {author} {\bibfnamefont {L.~S.}\ \bibnamefont
  {Theis}}, \bibinfo {author} {\bibfnamefont {F.}~\bibnamefont {Motzoi}}, \
  and\ \bibinfo {author} {\bibfnamefont {F.~K.}\ \bibnamefont {Wilhelm}},\
  }\href {http://dx.doi.org/10.1103/PhysRevA.93.012324} {\bibfield  {journal}
  {\bibinfo  {journal} {Phys. Rev. A}\ }\textbf {\bibinfo {volume} {93}},\
  \bibinfo {pages} {012324} (\bibinfo {year} {2016})}\BibitemShut {NoStop}%
\bibitem [{\citenamefont {Wang}\ \emph {et~al.}(2015)\citenamefont {Wang},
  \citenamefont {Zhang}, \citenamefont {Corcovilos}, \citenamefont {Kumar},\
  and\ \citenamefont {Weiss}}]{YWang2015}%
  \BibitemOpen
  \bibfield  {author} {\bibinfo {author} {\bibfnamefont {Y.}~\bibnamefont
  {Wang}}, \bibinfo {author} {\bibfnamefont {X.}~\bibnamefont {Zhang}},
  \bibinfo {author} {\bibfnamefont {T.~A.}\ \bibnamefont {Corcovilos}},
  \bibinfo {author} {\bibfnamefont {A.}~\bibnamefont {Kumar}}, \ and\ \bibinfo
  {author} {\bibfnamefont {D.~S.}\ \bibnamefont {Weiss}},\ }\href
  {http://dx.doi.org/10.1103/PhysRevLett.115.043003} {\bibfield  {journal}
  {\bibinfo  {journal} {Phys. Rev. Lett.}\ }\textbf {\bibinfo {volume} {115}},\
  \bibinfo {pages} {043003} (\bibinfo {year} {2015})}\BibitemShut {NoStop}%
\bibitem [{\citenamefont {Gilchrist}\ \emph {et~al.}(2005)\citenamefont
  {Gilchrist}, \citenamefont {Langford},\ and\ \citenamefont
  {Nielsen}}]{Gilchrist2005}%
  \BibitemOpen
  \bibfield  {author} {\bibinfo {author} {\bibfnamefont {A.}~\bibnamefont
  {Gilchrist}}, \bibinfo {author} {\bibfnamefont {N.~K.}\ \bibnamefont
  {Langford}}, \ and\ \bibinfo {author} {\bibfnamefont {M.~A.}\ \bibnamefont
  {Nielsen}},\ }\href
  {http://journals.aps.org/pra/abstract/10.1103/PhysRevA.71.062310} {\bibfield
  {journal} {\bibinfo  {journal} {Phys. Rev. A}\ }\textbf {\bibinfo {volume}
  {71}},\ \bibinfo {pages} {062310} (\bibinfo {year} {2005})}\BibitemShut
  {NoStop}%
\bibitem [{\citenamefont {Xia}\ \emph {et~al.}(2015)\citenamefont {Xia},
  \citenamefont {Lichtman}, \citenamefont {Maller}, \citenamefont {Carr},
  \citenamefont {Piotrowicz}, \citenamefont {Isenhower},\ and\ \citenamefont
  {Saffman}}]{Xia2015}%
  \BibitemOpen
  \bibfield  {author} {\bibinfo {author} {\bibfnamefont {T.}~\bibnamefont
  {Xia}}, \bibinfo {author} {\bibfnamefont {M.}~\bibnamefont {Lichtman}},
  \bibinfo {author} {\bibfnamefont {K.}~\bibnamefont {Maller}}, \bibinfo
  {author} {\bibfnamefont {A.~W.}\ \bibnamefont {Carr}}, \bibinfo {author}
  {\bibfnamefont {M.~J.}\ \bibnamefont {Piotrowicz}}, \bibinfo {author}
  {\bibfnamefont {L.}~\bibnamefont {Isenhower}}, \ and\ \bibinfo {author}
  {\bibfnamefont {M.}~\bibnamefont {Saffman}},\ }\href
  {http://dx.doi.org/10.1103/PhysRevLett.114.100503} {\bibfield  {journal}
  {\bibinfo  {journal} {Phys. Rev. Lett.}\ }\textbf {\bibinfo {volume} {114}},\
  \bibinfo {pages} {100503} (\bibinfo {year} {2015})}\BibitemShut {NoStop}%
\end{thebibliography}%
\bibliographystyle{apsrev4-1}

\end{document}